# Thickness-dependent crack suppression and wrinkle formation in freestanding $BaTiO_3$ membranes

*Rajesh Mandal**, *Ajay Kumar, Nini Pryds**

*Department of Energy Conversion and Storage, Technical University of Denmark (DTU), Fysikvej, 310, 2800 Kgs. Lyngby, Denmark*

Email for correspondence: ***rajema@dtu.dk*** **,** ***nipr@dtu.dk***

## Abstract

*Freestanding ferroelectric oxide membranes offer a promising route toward silicon-integrated low-power electronic and memory devices, but reproducible membrane quality remains challenging, particularly in ultrathin $BaTiO_3$ (BTO), where cracking and wrinkling hinder integration. Here, we show that crack and wrinkle formation in released $BaTiO_3$ membranes can be tuned by jointly controlling oxide thickness and polymer support thickness. Reducing the $BaTiO_3$ thickness from 15 nm to 5 nm suppresses large-area, optically visible cracking across the analysed regions, but promotes dense nanoscale wrinkling, revealing a trade-off between fracture mitigation and morphological instability. In 10 nm $BaTiO_3$ membranes, polymer support thickness further modulates wrinkling, with an intermediate thickness reducing wrinkle density. Piezoresponse force microscopy reveals time-dependent evolution of written contrast after release, indicating that apparent ferroelectric response is influenced by polarization switching, post-release morphology, interfacial contact, charge screening, and strain relaxation. These results establish processing guidelines for integrating freestanding ferroelectrics into future devices.*

## Introduction

**I**ntegration of high-quality ferroelectric oxides with a Si-based platform is essential for the development of oxide-based ferroelectric RAM, memristors, and next-generation low-power computational devices [1-3]. The recent development of freestanding oxide membranes has enabled new routes for further advancement in this direction. [4] Oxide membranes released from epitaxial strain imposed by the parent substrate can retain their material properties and crystalline quality, both of which are essential for robust device fabrication. In the past few years, significant research has been conducted to explore various growth and release mechanisms to achieve large-area, defect-free, flat membranes [5-8]. Despite progress in membrane transfer and in integrating membranes with silicon, broader applicability remains limited by a lack of precise control and reproducibility of membrane quality. [9]. Ferroelectric oxide membranes are particularly important in this context. For the prototypical lead-free, room-temperature ferroelectric barium titanate [10–12], numerous studies have reported extensive structural characterization and modulation of ferroelectric ordering toward integration with silicon-based, CMOS-compatible platforms [13–16]. Although these results are promising, membrane quality remains a major challenge. Two of the most challenging issues are the formation of cracks and wrinkles [17,18], and the use of $BaTiO_3$ (BTO) remains, therefore, particularly challenging in this context. Periodic wrinkle patterns have often been observed in thin BTO membranes, and these are intrinsic to the membrane property [19]. Though wrinkles are undesirable and sometimes uncontrollable, they have been found to modulate the microscopic polar ordering in membranes [20]. Generally, wrinkles appear as soon as the membrane is released from the epitaxial strain of the parent substrate. Ferroelastic BTO membranes attached to soft polymers produce a wrinkled pattern after release. After release, the soft polymer tries to recover its original length, imposing compression on the much stiffer BTO membrane. The BTO relieves that compression by out-of-plane buckling, and because BTO is ferroelastic, its domain variants further reorganize to accommodate the spatially varying strain, stabilizing periodic, zigzag, or labyrinth wrinkle patterns. The differently strained wrinkle regions induce either *in-plane a* domain (tensile strain) or *out-of-plane c* domains (compressive strain) [13,21]. It is also common for the BTO membrane to undergo spontaneous bending after being released from epitaxial strain, which leads to the formation of folded rolls with a rearrangement of microscopic ferroelectric-ferroelastic domains [22]. The formation of cracks in the membrane is another major concern for successful integration into the devices. Only a very limited number of ongoing efforts focus on controlling crack formation and understanding the key parameters that drive it. [23-28] Along with the growth parameters of the epitaxial layers, the lattice parameter of the sacrificial layer, the layer thickness (both membrane & sacrificial layer) [29] and the mechanical property of the support polymer [30] also play a crucial role in determining the stiffness of the membrane. In most cases, cracks form due to the sudden release of stress gradients caused by lattice mismatch between the sacrificial layer and the desired oxide. A comprehensive study by Yun et al. has shown that gradual tuning of the $Sr_3Al_2O_6$ (SAO) sacrificial layer's lattice parameter via doping can make a $SrTiO_3$ membrane perfect [31]. An effective way to adjust the epitaxial mismatch strain after membrane release, as demonstrated by Moon et al., is to engineer an elastically graded polymer for the support layer. [30] The non-centrosymmetric tetragonal lattice in BTO makes it even more challenging to fabricate a large-area membrane without optically visible large-area cracks in the analyzed regions, especially in the ultra-thin regime, which is necessary for practical devices. In summary, although major progress has been made in transferring freestanding oxide membranes and suppressing fracture through sacrificial-layer and polymer engineering, reproducible control of crack and wrinkle formation in ultrathin ferroelectric $BaTiO_3$ remains limited. In particular, the relative roles of membrane thickness and polymer-support conditions in determining whether strain is relieved by cracking or wrinkling remain insufficiently understood in the ultrathin regime relevant to device integration. A systematic comparison under fixed growth and transfer conditions is therefore needed to distinguish thickness-dependent trends from variations introduced by the release and transfer process. Here, we address these points by comparing freestanding BTO membranes with thicknesses ranging from 5 to 15 nm and by varying the polymer support thickness. The results provide an empirical processing map for reducing visible cracking while identifying wrinkling and time-dependent polarization response as remaining limitations.

## Wrinkle-to-Crack transition in ultrathin freestanding $BaTiO_3$ membranes

To examine how membrane thickness affects post-transfer morphology, we prepared freestanding BTO membranes with thicknesses ranging from 5 nm to 15 nm, using SAO as a sacrificial layer and CAB as the transfer support (Detailed growth conditions are provided in the Methods section). Across this thickness range, we observe a clear evolution in the dominant strain-relief mode: thicker membranes exhibit visible crack formation, whereas thinner membranes remain apparent crack-free over larger areas but develop increasingly dense wrinkles. Because the growth and transfer conditions were kept nominally constant while the nominal BTO thickness was varied, this trend points to a thickness-dependent post-release mechanical response of the membrane in the ultrathin regime. However, thickness can also influence residual strain, defect density, adhesion, and release kinetics; therefore, the observed trend should be interpreted as an empirical thickness dependence under the present transfer protocol rather than as a universal thickness-only effect. This is illustrated schematically in **Fig. 1**.

**Fig. 2 (a) – (d)** shows dark-field optical images and corresponding AFM topography for membranes with thicknesses of 5, 10, 12 and 15 nm. Here, a constant CAB thickness of 200 $\mu m$ is used as the support layer. The 12 nm and 15 nm membranes exhibit clearly visible cracks over large areas, whereas no visible large-area cracks are observed for the 5 nm and 10 nm membranes. In contrast, AFM reveals that wrinkle formation becomes increasingly pronounced as the membrane thickness decreases. The 15 nm-thick membrane exhibits a smooth surface between cracks, whereas the 12 nm-thick membrane shows the onset of wrinkles.

In contrast, the 5 nm and 10 nm membranes exhibit a higher density of nanoscale wrinkling than the 12 nm and 15 nm membranes. The crack and wrinkle densities are quantified in **Fig.2 (e-f)**. These observations indicate a thickness-dependent crossover in post-release morphology, from crack-dominated behavior in thicker membranes to wrinkle-dominated strain accommodation in thinner membranes. The crossover thickness lies between 10-12 nm and is visible from **Fig. 1 (b)** and **(c)**. Corresponding bright-field large images are presented in Figure **SI-2** in the Supplementary Information.

The observed trend is consistent with a thickness-dependent shift in the elastic properties and mechanisms that accommodate released epitaxial strain, in which thinner membranes relieve stress preferentially through wrinkling rather than fracture. In **Fig. 2 (e)** and **(f)** we plot the average crack and wrinkle density with the membrane thickness. Crack density is based on optically visible cracks in dark-field images, whereas wrinkle density is extracted from AFM topography. Therefore, the two metrics probe different length scales: optical imaging captures large-area fracture features, while AFM captures local nanoscale out-of-plane deformation. The absence of optically visible cracks should therefore not be interpreted as the absence of all defects, but rather as suppression of large-area cracks within the analyzed fields of view. Interestingly, this trend differs from earlier reports, which found that thinner BTO membranes (8.33 nm) were more prone to cracking than thicker ones (16.7 nm) [29]. This discrepancy likely reflects differences in the interconnected physical properties of the membrane, sacrificial layer, transfer support, and release conditions. Our results, therefore, suggest that crack formation in freestanding BTO cannot be described by thickness alone; rather, it depends on the detail growth and membrane fabrication processing *t*, particularly in the ultrathin regime.

After identifying the 10 nm $BaTiO_3$ as the approximate thickness for which optically visible large-area cracking is substantially reduced under the present transfer conditions, we examined whether the CAB support thickness affects the remaining wrinkle morphology. The CAB thickness was varied from 200 µm to 500 µm, while the $BaTiO_3$ thickness was kept constant at 10 nm. As shown in **Fig. 3 (a–d)**, no optically visible large-area cracks are observed in the analyzed regions for any of the CAB thicknesses. However, the AFM topography shows substantial changes in wrinkle density and wrinkle height. Intermediate CAB thicknesses of 300–400 µm yield lower wrinkle density and lower wrinkle height than the 200 µm and 500 µm cases. The representative AFM topography images suggest that intermediate CAB thicknesses of 300–400 µm exhibit less pronounced wrinkling than the 200 µm and 500 µm cases. However, additional quantitative analysis would be required to establish whether this apparent non-monotonic dependence is statistically robust. This non-monotonic behavior indicates that the transfer support influences wrinkle formation but cannot be described by CAB thickness alone. Other parameters, including polymer modulus, residual stress, solvent drying, adhesion, and the receiving-substrate surface, are also likely to contribute. Under the tested conditions, 10-nm-thick $BaTiO_3$ membranes did not exhibit large-area, optically visible cracking across CAB support thicknesses between 200 µm and 500 µm. This observation suggests that once the membrane thickness is reduced to 10 nm, visible crack formation is less sensitive to CAB thickness than is wrinkle morphology. However, because the polymer modulus and adhesion were not varied independently, the present results do not establish independence of the results from support-layer stiffness. Similarly, Moon et al. achieved a large-area, crack-free BTO membrane using an elastically graded polymer (EGP) support layer with a thickness of 200-300 nm for a 10 nm-thick BTO membrane [30]. The formation of wrinkles with an average height of 18 nm to 30 nm, observed in our cases on thin membranes, appears to depend on membrane elasticity, the transfer method, and the substrate surface. We observe changes in the wrinkle pattern depending on the surface to which the transferred membrane is attached, as shown in Figure **SI-4,** where the BTO membrane has been transferred onto gold-coated Si and PDMS.

The observed change from crack- to wrinkle-dominated morphology can be qualitatively understood as a competition between fracture and out-of-plane deformation during strain relaxation. For a thin elastic plate, the bending stiffness scales approximately as $D \propto Et^3$, where $E$ is the elastic modulus and $t$ is the membrane thickness. Therefore, reducing the $BaTiO_3$ thickness significantly lowers the energetic cost of bending and may enable the membrane to accommodate strain via nanoscale wrinkling rather than crack propagation. In contrast, thicker membranes have higher bending stiffness and may be less able to relax through out-of-plane deformation, making fracture a more favorable relaxation pathway. This interpretation is qualitative because residual strain, fracture toughness, adhesion energy, support-layer modulus, and transfer-induced stresses were not independently measured in the current study.

## Temporal electromechanical stability in released $BaTiO_3$

To evaluate the local electromechanical response before and after membrane release, we performed vertical PFM measurements on substrate-clamped and released $BaTiO_3$ films. **Fig. 4(a)** shows the topography, PFM phase, and PFM amplitude of the epitaxial $BaTiO_3$ film on Nb:$SrTiO_3$ after writing oppositely poled regions using +10 V and −10 V tip biases. The clear phase contrast between the written regions is consistent with switchable electromechanical response in the substrate-clamped film.

After release and transfer onto Pt-coated Si, PFM measurements were performed on a 15 nm $BaTiO_3$ membrane region that had good local contact with the bottom electrode. This thickness was selected because it yielded a stronger, more readily detectable PFM signal under the present measurement conditions. However, because the 15 nm membranes are also more susceptible to visible cracking than the thinner 5–10 nm membranes, the released-membrane PFM results should not be taken as representative of the optimized crack-suppressed morphology. As shown in **Fig. 4(b)**, voltage writing produces a spatially fragmented PFM phase and amplitude response rather than a uniform written domain pattern. The reduced uniformity compared with the clamped film suggests that the released membrane geometry, local strain relaxation, cracking, and nonuniform electrode contact influence the measured electromechanical response.

To examine the temporal stability of the written contrast, the same region was imaged repeatedly at 8 min intervals after writing. The initially distinguishable phase contrast between the +10 V and −10 V written regions decrease with time, and the amplitude contrast also becomes weaker. Quantitative analysis in **Fig. 4(d)** shows a gradual evolution from an initially bipolar phase-contrast toward a more unipolar response, with the amplitude and phase ratio approaching 1 in 56 min.

The absence of comparable temporal decay in the substrate-clamped sample, together with the pronounced decay observed in the released membrane, suggests that the post-release boundary conditions strongly influence the measured PFM response. In the clamped film, the substrate provides mechanical constraint and an efficient charge-dissipation pathway. After release, the membrane becomes weakly constrained, and its electrical contact to the receiving substrate can be spatially nonuniform. During ± 10 V tip biasing, charge injection and trapping at surface states, oxygen-vacancy-related defects, adsorbates, and membrane interfaces may contribute to the measured contrast [32,33]. Because the released membrane geometry can provide poorer grounding and reduced screening compared with the clamped film, the written PFM contrast may contain electrostatic component immediately after poling.

The gradual transformation from initially bipolar phase contrast toward weaker and more unipolar contrast, therefore, cannot be assigned solely to stable remanent ferroelectric polarization. Stable ferroelectric switching would be expected to preserve a more persistent phase difference between oppositely poled regions after the writing bias is removed. In contrast, charge-mediated PFM contrast can decay as trapped charges redistribute, leak, or become screened by ambient adsorbates. The simultaneous reduction in phase and amplitude contrast is therefore consistent with a mixed electromechanical and electrostatic response in the released membrane. Recent studies on freestanding oxide membranes have highlighted that substrate removal fundamentally alters dielectric screening and charge dissipation pathways, making suspended structures more susceptible to long-lived electrostatic effects and charge-relaxation phenomena [34]. Therefore, achieving smoother membranes with reduced wrinkles and cracks, together with an optimized interface that enables efficient charge dissipation while preserving polarization stability, will be essential for realizing reliable intrinsic ferroelectric switching in freestanding oxide membranes. Further measurements, including switching-spectroscopy PFM, controlled-atmosphere PFM, Kelvin probe or electrostatic force microscopy, and independent electrical hysteresis measurements, will be required to distinguish intrinsic ferroelectric switching from charge-mediated contrast in released $BaTiO_3$ membranes.

## Conclusion

In summary, we investigated the post-transfer morphology of freestanding $BaTiO_3$ membranes as a function of membrane thickness and CAB support thickness. Under the present growth and transfer conditions, reducing the $BaTiO_3$ thickness, e.g., from 15 nm to 5 nm, suppresses optically visible large-area cracking within the analyzed regions, while nanoscale wrinkling becomes increasingly pronounced. This behavior indicates a transition from crack-dominated to wrinkle-dominated strain relaxation within the tested thickness range. For 10 nm $BaTiO_3$ membranes, varying the CAB support thickness modifies wrinkle morphology without reintroducing optically visible large-area cracking in the analyzed regions, demonstrating that support-layer conditions remain important even after visible crack suppression. However, the present results do not establish a universal thickness threshold for crack suppression because residual strain, adhesion energy, polymer modulus, drying stress, and transfer-induced stresses were not independently varied.

PFM measurements further show that written contrast in released $BaTiO_3$ membranes can evolve substantially over time, suggesting that post-release morphology, local strain relaxation, electrical contact, and charge-screening conditions influence the apparent nanoscale response. Additional electrostatic and electrical controls will be required to separate stable ferroelectric switching from charge-mediated PFM contrast in released membranes. These results identify practical transfer conditions for reducing visible cracking in ultrathin $BaTiO_3$ membranes, while also showing that wrinkle suppression, interface engineering, and polarization stability remain unresolved challenges for device integration [35].

## Methods

**Thin Film Growth.** The BTO films, 15 *nm* to 5 *nm* thick, were grown on a treated $TiO_2$-terminated $SrTiO_3$ (STO) (001)-oriented substrate with a 10 nm sacrificial layer of $Sr_3Al_2O_6$ (SAO) using high-energy electron diffraction (RHEED) - assisted pulsed laser deposition (PLD) (Coherent KrF UV laser, wavelength 248 nm). The SAO layer was grown at a temperature of 750 °C with a laser fluence of 1.2 $J/cm^2$ and an $O_2$ partial pressure of $2\times10^{-5}$ mbar. The BTO film was grown on SAO at 650° C with a laser fluence of 1 $J/cm^2$ and $O_2$ partial pressure of $5\times10^{-3}$ mbar. Both BTO and SAO are grown with a pulse repetition rate of 2 Hz with a chamber base pressure of $1\times10^{-8}$ mbar. The STO/SAO/BTO stack is post annealed after the deposition at 500° C at 200 mbar for 30 mins and then cooled down to room temperature at 10° C/min. Crack density was quantified from dark-field optical images by measuring the total crack length per unit area using ImageJ. Wrinkle density and wrinkle height were extracted from AFM topography over multiple scan areas for each sample condition. For each thickness, at least 5 membranes and 10 image regions were analyzed, and the plotted values represent mean ± standard deviation.

**Membrane release and transfer.** After growth, a cellulose acetate butyrate (CAB) support layer was applied to the BTO/SAO/STO heterostructure before release. CAB layers with nominal thicknesses of 200, 300, 400, and 500 µm were prepared under identical casting and drying conditions, and their thicknesses were measured at multiple positions before release. Unless otherwise stated, the CAB thickness was fixed at 200 µm for the BTO-thickness-dependent study. The CAB-supported heterostructures were immersed in deionized water to selectively dissolve the $Sr_3Al_2O_6$ sacrificial layer. The release was allowed to proceed until the CAB/BTO membrane was fully separated from the STO substrate. The released membranes were then rinsed in deionized water and transferred onto the target substrates by scooping. After transfer, the samples were dried under identical ambient conditions before optical microscopy, AFM, and PFM measurements. More illustrated transfer processes can be found in supplementary reference [36].

**Quantification of crack and wrinkle morphology.** Crack density was quantified from dark-field optical images as the total visible crack length divided by the analyzed image area of $100 \times 100\ \mu m^2$.

Wrinkle density was quantified from AFM topography images as the total wrinkle-ridge length divided by the analyzed scan area of $10 \times 10\ \mu m^2$. Wrinkle height was extracted from AFM line profiles as the height difference between the wrinkle ridge and the adjacent local membrane surface. For each sample condition, at least five independently transferred membranes and ten image regions were analyzed. The values plotted in Figures 2 and 3 represent mean ± standard deviation. All images were analyzed using the same thresholding and segmentation criteria in ImageJ to ensure consistent comparison between thicknesses and CAB-support conditions. Because the analysis is based on finite fields of view, the reported densities should be interpreted as local morphology metrics for the analyzed regions rather than wafer-scale defect densities.

**Characterization.** Large images are taken with an optical microscope with magnification up to 100x and dark-field mode with constant exposure of 200 ms**.** AFM and PFM imaging have been done with Bruker Dimension Icon in contact mode using Bruker conducting probe SCM-PIT-V2 with spring constant of 3N/m at a vertical contact resonance frequency of 280 kHz. XRD is carried out with Rigaku Smartlab.

## Acknowledgements

The authors acknowledge the support from the ERC Advanced (NEXUS, grant no. 101054572), research grant (VIL73726) EURIKA from Villum Fonden and ERC Synergy Grant METRIQS (no. 101167432).

### Author Contributions

RM and NP conceived, supervised the concept and wrote the manuscript. RM prepared and characterized the membranes as well as performed the microscopy and PFM measurements. AK helped in writing the manuscript, data analysis and PFM measurements. All authors discussed the results and revised the manuscript.

## Data availability

The data that supports the findings of this study are available from the corresponding authors upon reasonable request.

## Declarations

**Conflict of interest** The authors declare no competing financial interest or conflict.

## Figures with Captions

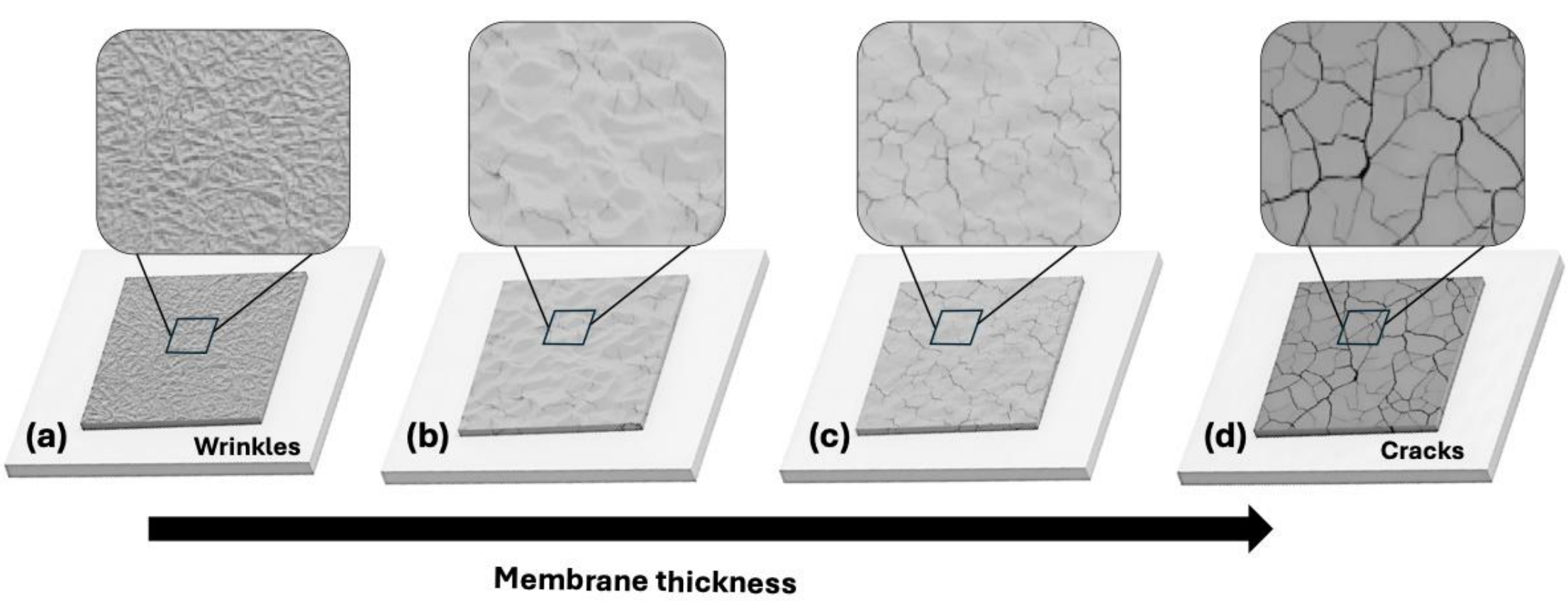

**Figure 1: Schematic illustration of thickness-dependent strain relaxation in freestanding $BaTiO_3$ membranes.** Thicker membranes relax transfer-induced strain primarily through optically visible crack formation (d), whereas thinner membranes (a) more readily accommodate strain through out-of-plane wrinkling. The schematic illustrates the proposed crack-to-wrinkle transition (a to d) observed within the tested thickness range.

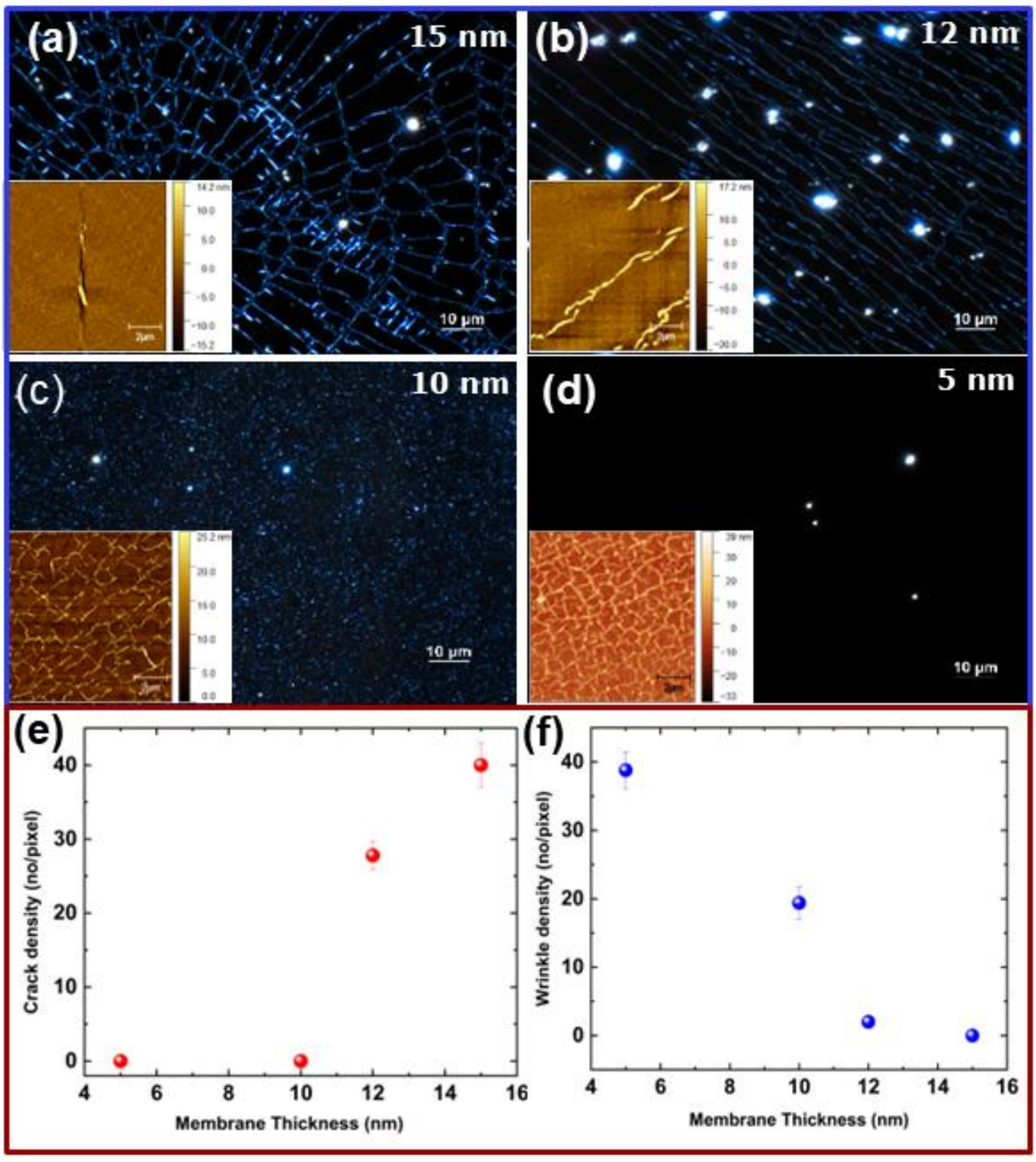

**Figure 2: Thickness-dependent evolution of crack and wrinkle formation in freestanding BTO membranes.** Dark-field optical images and corresponding AFM topography of transferred BTO membranes with thicknesses of (a) 15 nm, (b) 12 nm, (c) 10 nm, and (d) 5 nm. Visible large-area cracks are observed in the 15 nm and 12 nm membranes, whereas no optically visible large-area cracks are observed in the 10 nm and 5 nm membranes within the imaged regions. These thinner membranes instead exhibit increasingly dense nanoscale wrinkling. Quantification of (e) crack density and (f) wrinkle density reveals a crossover from crack-dominated to wrinkle-dominated morphology with decreasing thickness.

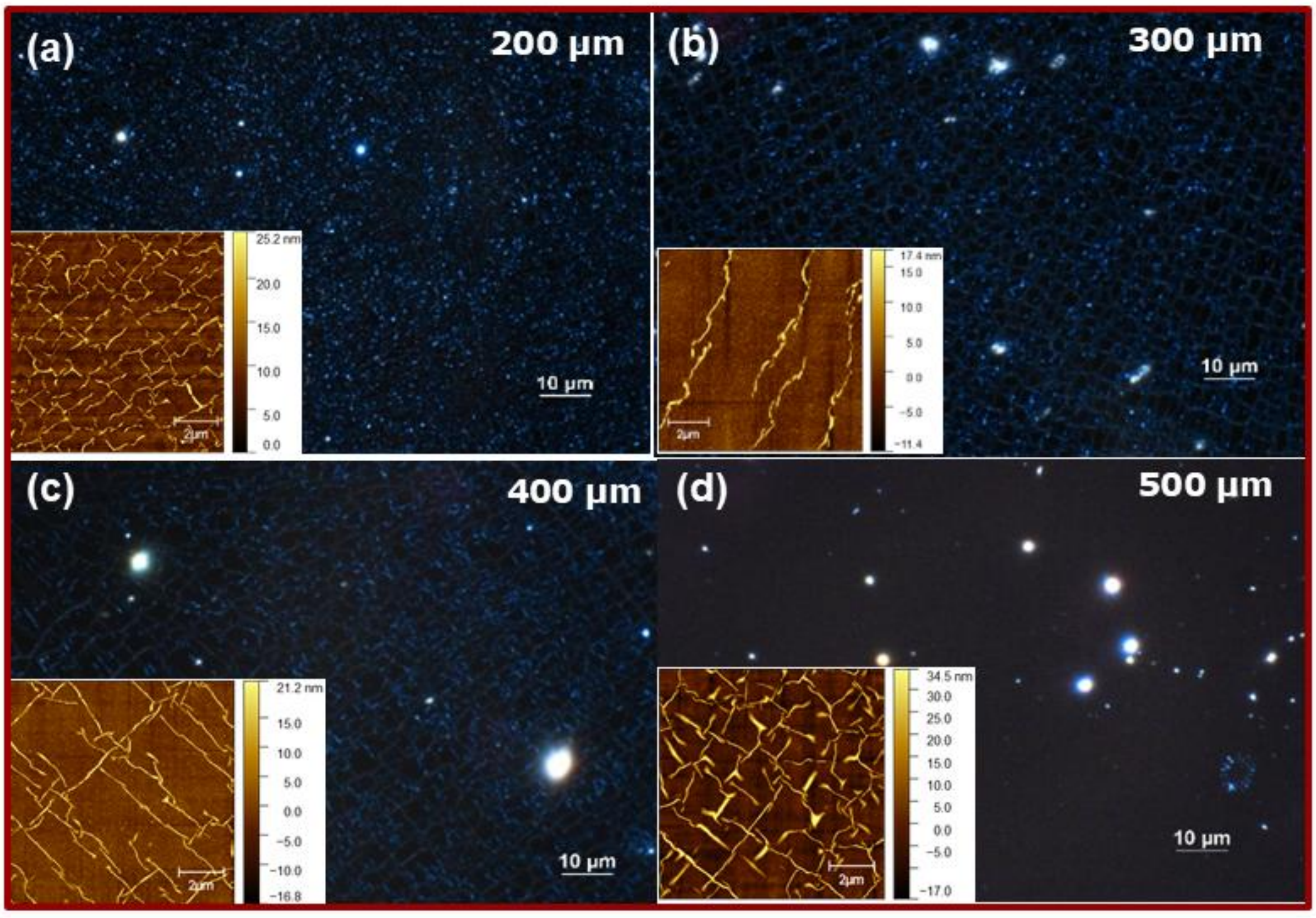

**Figure 3: Influence of CAB support thickness on wrinkle morphology in 10 nm freestanding BTO membranes.** Dark-field optical images and AFM topography of 10 nm BTO membranes transferred using CAB thicknesses of (a) 200 µm, (b) 300 µm, (c) 400 µm, and (d) 500 µm. No large-area cracks are visible under any condition, but the wrinkle morphology varies significantly with CAB thickness. Intermediate CAB thicknesses (300 and 400 µm ) yield visibly less wrinkles with lower wrinkle height as seen from the AFM topography and the scalebar, indicating a non-monotonic dependence of wrinkle formation on transfer support mechanics. The representative AFM images suggest the wrinkle formation is sensitive to the thickness of the transfer-support polymer.

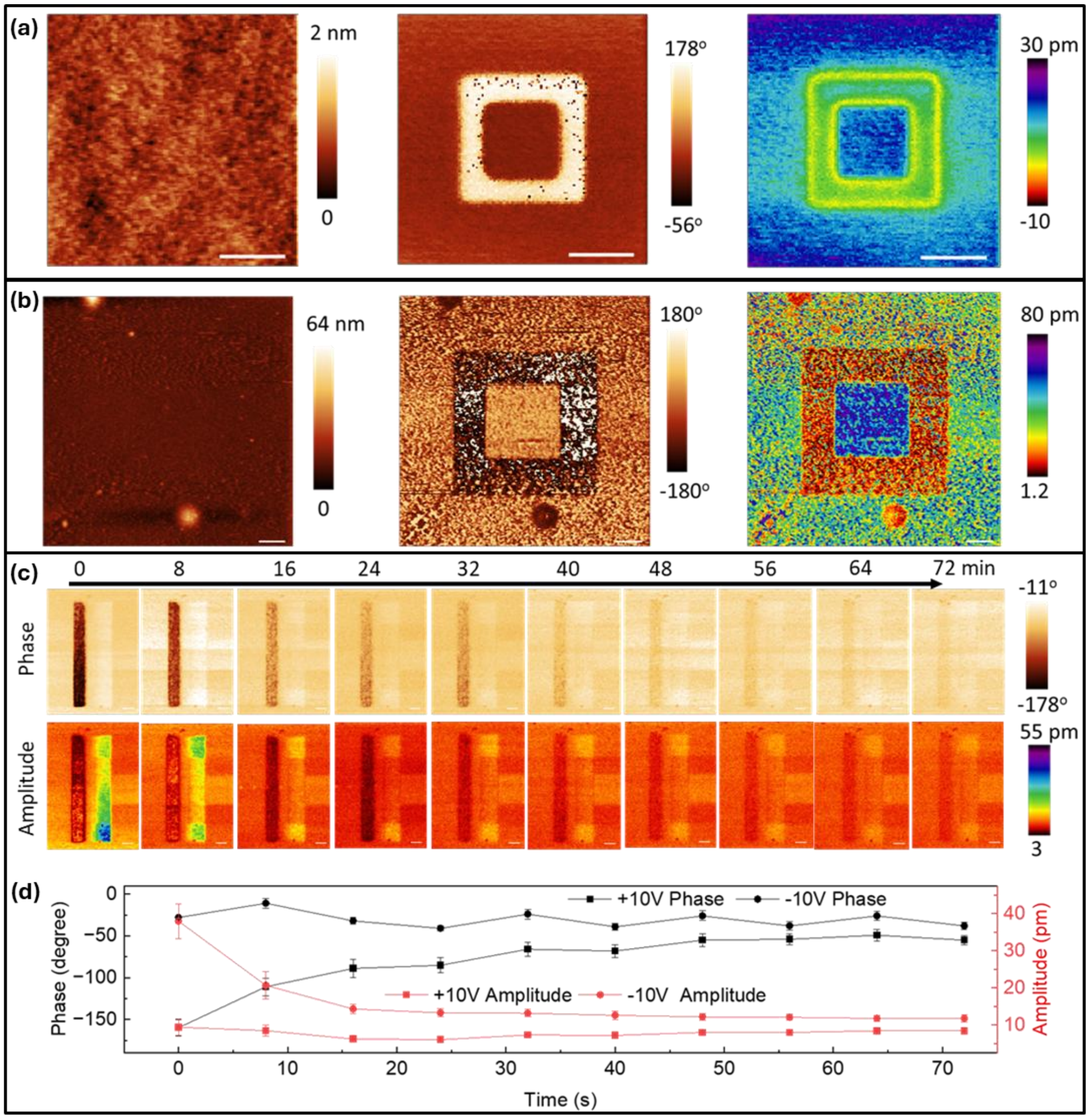

**Figure 4: Time-dependent PFM response of clamped and released $BaTiO_3$ membranes.** (a) AFM topography and corresponding vertical PFM phase and amplitude images of the epitaxial $BaTiO_3$ film on Nb:$SrTiO_3$ after writing oppositely poled regions using +10 V and −10 V tip biases. (b) AFM topography and PFM response of a released 15 nm $BaTiO_3$ membrane transferred onto Pt-coated Si after voltage writing. (c) Time-dependent PFM phase and amplitude images collected at 8 min intervals from the written region over 72 min. (d) Quantitative analysis of the phase and amplitude evolution, showing a gradual reduction of the initially written contrast over the measurement period. The observed relaxation may arise from intrinsic polarization relaxation, strain relaxation, charge screening, or nonuniform electrical contact.

## Supplementary Information

# Thickness-dependent crack suppression and wrinkle formation in freestanding $BaTiO_3$ membranes

*Rajesh Mandal*, Ajay Kumar, Nini Pryds**

*Department of Energy Conversion and Storage, Technical University of Denmark (DTU), Fysikvej, 310, 2800 Kgs. Lyngby, Denmark*

Email for correspondence: rajema@dtu.dk , nipr@dtu.dk

**Supplementary Figures:**

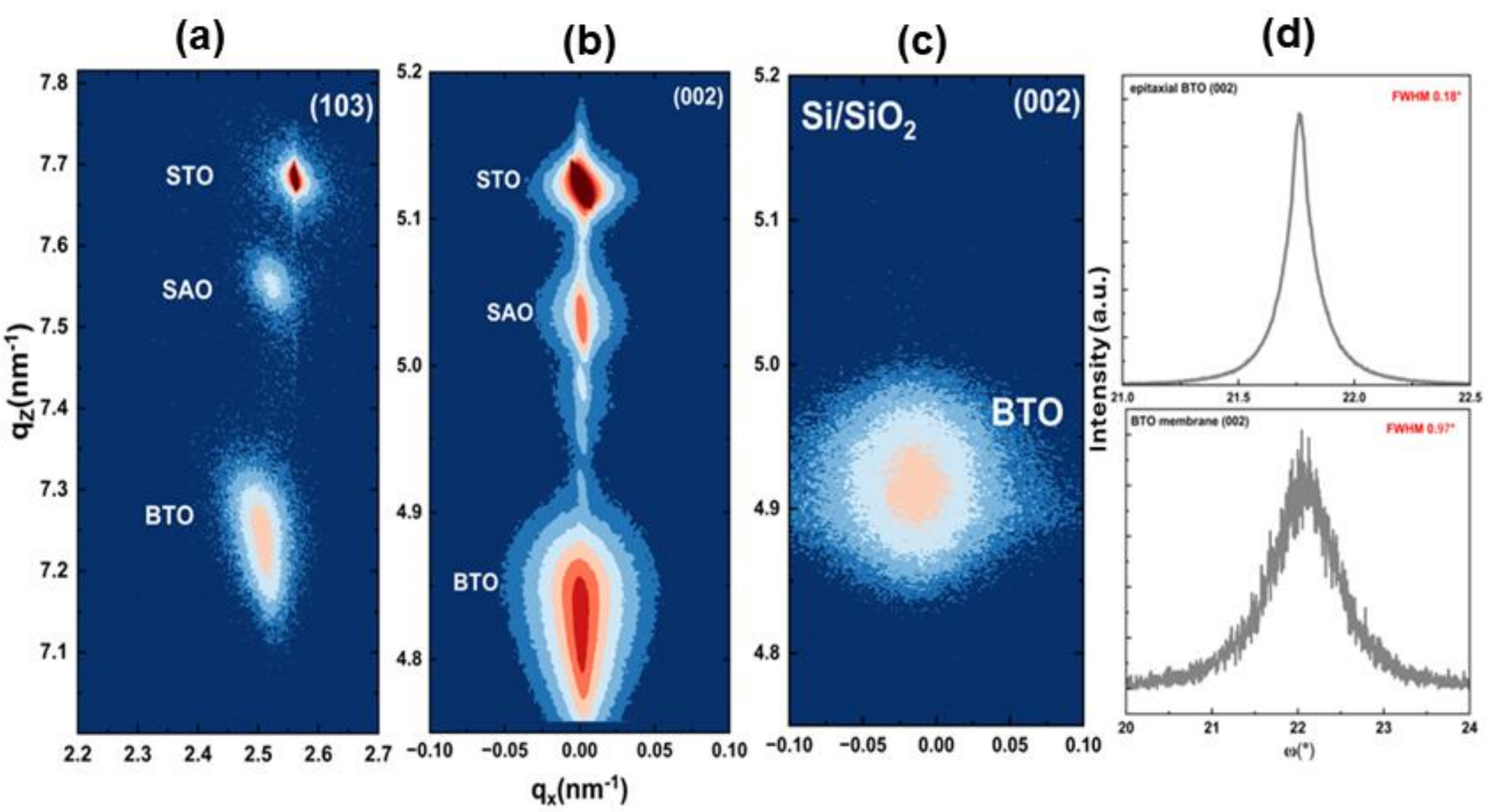

**SI 1:** Reciprocal Space Mapping (RSM) of (a) & (b) epitaxial BTO/SAO layer on STO for [103] and [002] planes, (c) BTO membrane transferred on $Si/SiO_2$. (d) Rocking curve FWHM for the epitaxial BTO layer as well as the BTO membrane.

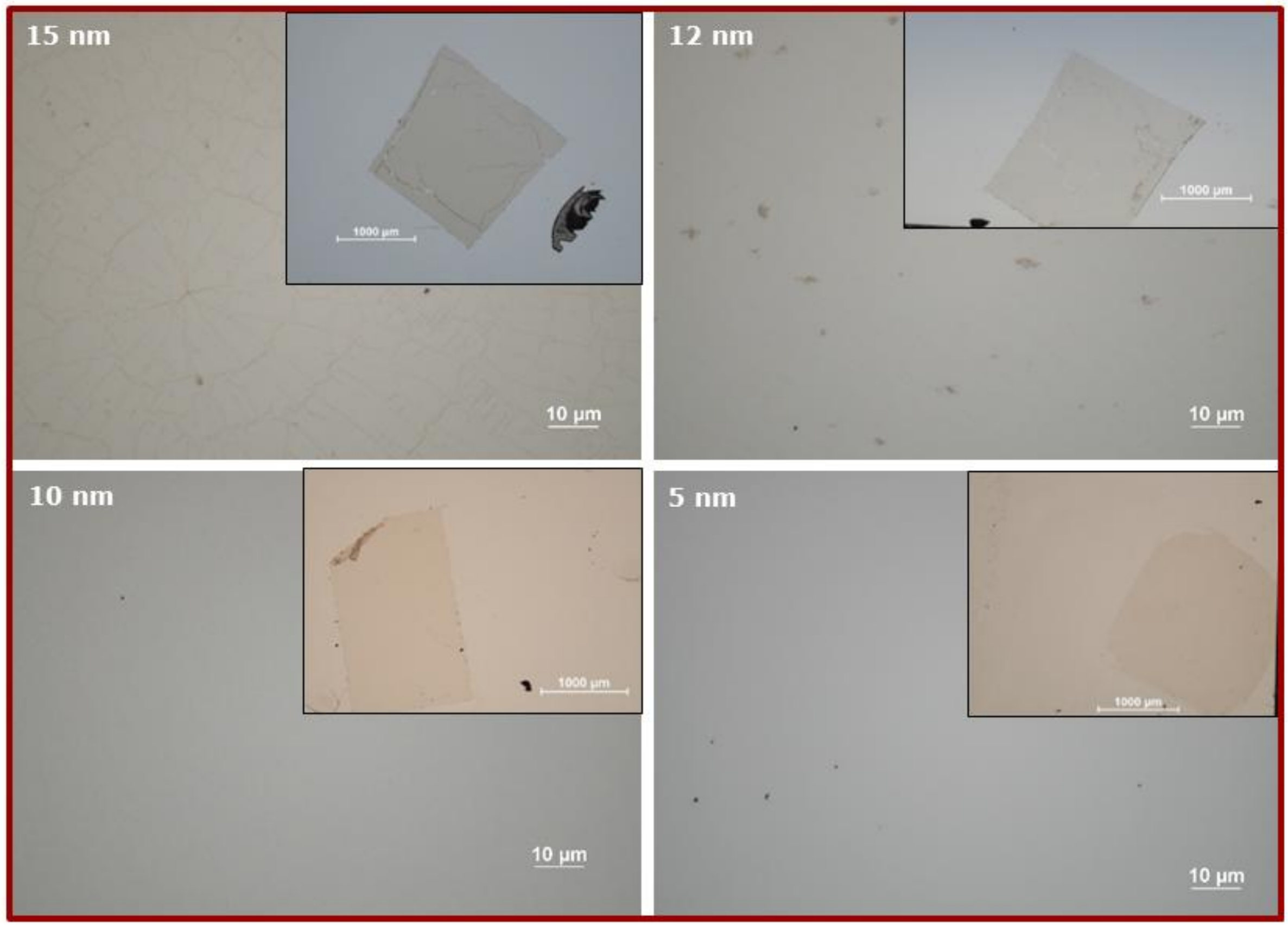

**SI 2:** Bright field large area image of the BTO membranes for the thickness ranges from 15 *nm* to 5 *nm*.

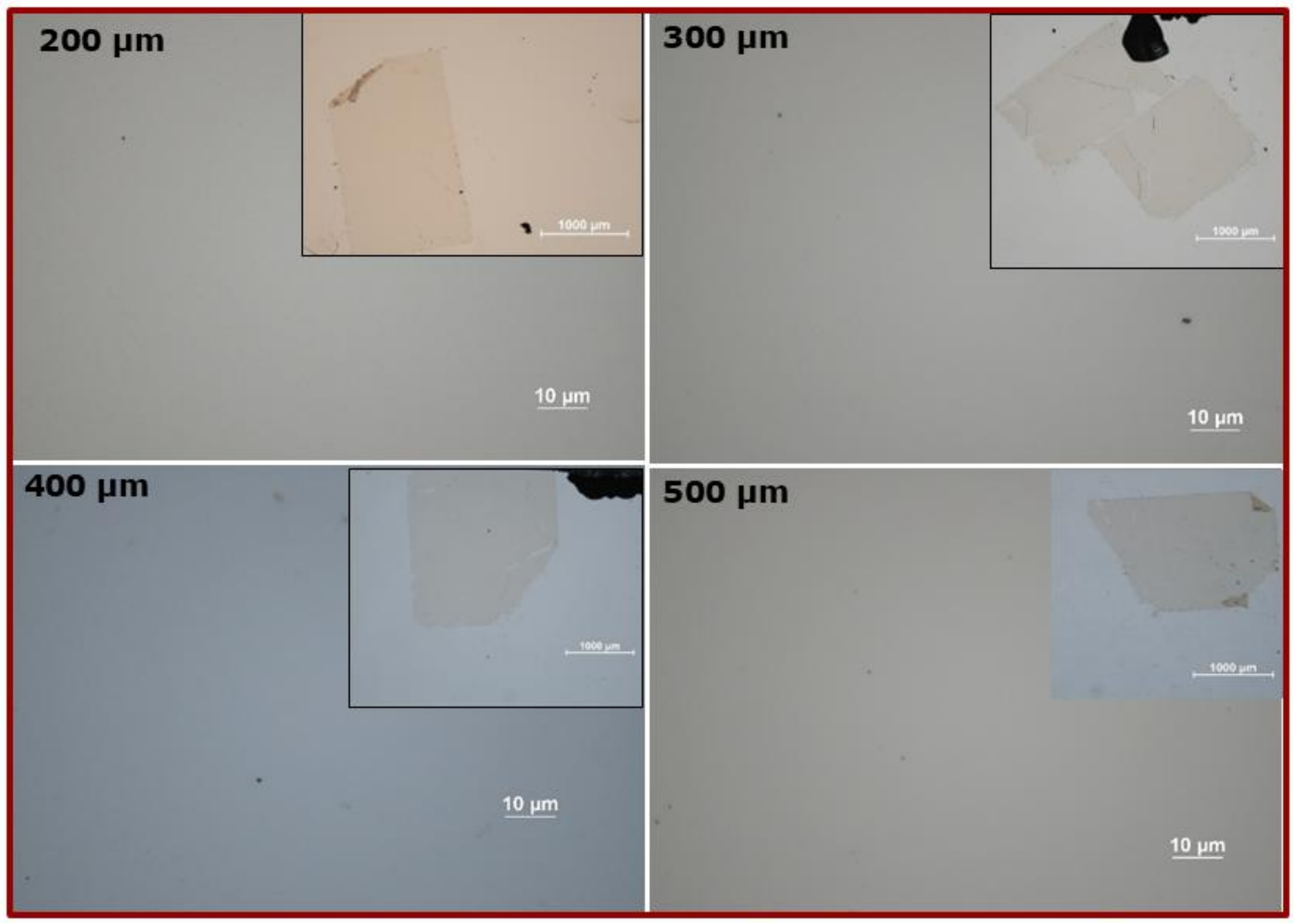

**SI 3:** Bright field large area image of the BTO membranes for the support polymer CAB thickness ranges from 200 µm to 500 µm.

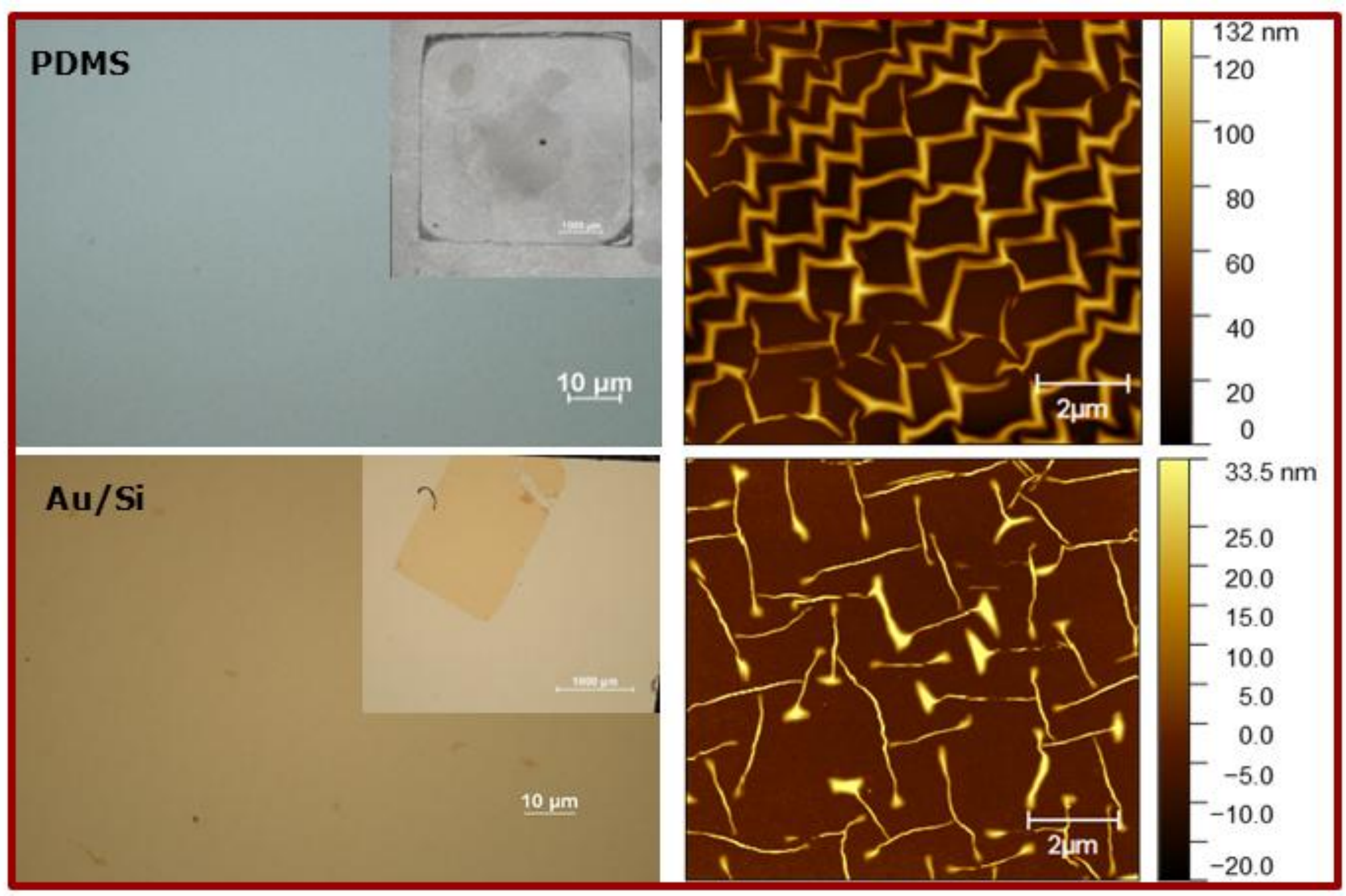

**SI 4:** Bright field large area image as well as the AFM topography of the BTO membranes transferred on PDMS and gold coated silicon substrate.

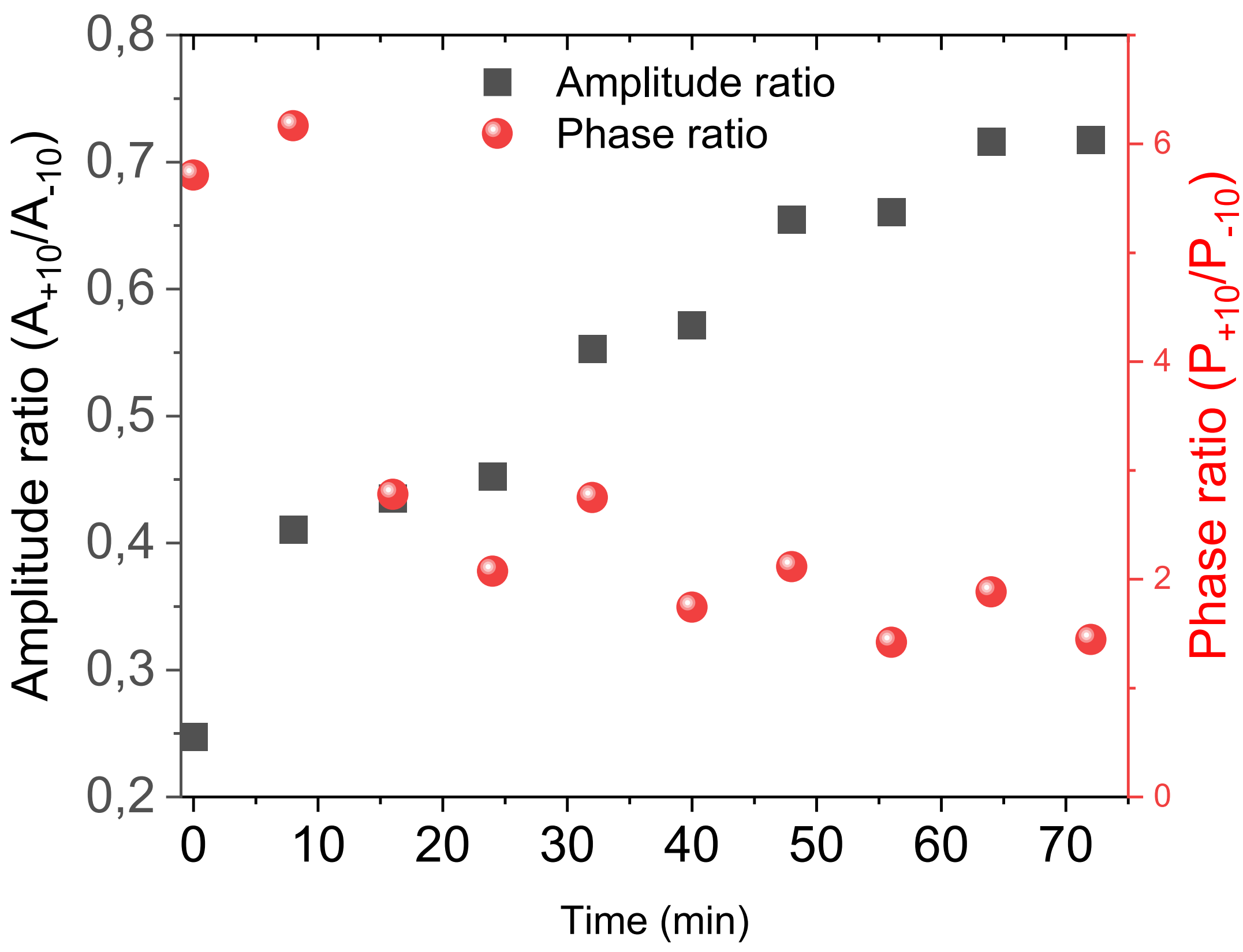

**SI 5:** Time evolution of the contrast ratio of the Amplitude and Phase for ±10 V